\documentclass[12pt]{article}
\input epsf
\usepackage{amssymb}
\usepackage{amsmath}
\usepackage{graphicx}
\makeatletter
\@addtoreset{equation}{section}
\makeatother

\def\be{\begin{equation}}
\def\ee{\end{equation}}

\def\ba{\begin{array}}
\def\ea{\end{array}}

\newcommand{\bea}{\begin{eqnarray}}
\newcommand{\eea}{\end{eqnarray}}
\def\N{$\cal N$}
\def\E {$E_{(7,7)}$}

\begin{document}
\hfill{}

\begin{flushright}
SU-ITP-2008-30
\end{flushright}

\vskip 1cm

\vspace{15pt}

\begin{center}
{ \LARGE {\bf  \N=8  Supergravity   4-point Amplitudes  \\
\vskip 0.8cm

  }}

\vspace{1.5cm}

  {\large \bf  Renata Kallosh, Ching Hua Lee and Tomas Rube
 } \\[7mm]

{

\ Department of Physics, Stanford University,  Stanford
 CA 94305-4060, USA}

\vspace{10pt}

    \vspace{15pt}

\vspace{10pt}

\end{center}

\begin{abstract}
We present the  explicit expressions in \N=8 supergravity for the bosonic 4-particle tree and 1-loop amplitudes including vectors and scalars.    We also present the candidate  4-point UV divergences  in a form of  helicity amplitudes, corresponding to 3-loop manifestly \N=8 supersymmetric and Lorentz covariant counterterm.  This may shed some light on the 3-loop  finiteness of \N=8 SG  and on a conjectured higher loop finiteness.

  We  perform a supersymmetric deformation to complex momentum of the 4-point generating function including  higher-loop counterterms and the 1-loop UV finite amplitudes.  Using the explicit form of the scalar part of the 3-loop counterterm and of the  1-loop UV finite scalar 4-point amplitudes we find that they both
    have an unbroken  \E  \, symmetry. We derive from  \E \,  symmetry the  low-energy theorem  for the 1-loop $n$-point amplitudes.

\end{abstract}

\newpage


\section{Introduction}

Recently it was shown that \N=8 SG \cite{Cremmer:1979up} is UV finite at the 3-loop level   \cite{Bern:2007hh}.  This is in agreement with the expectation from the unitarity cut method for computing amplitudes in maximally supersymmetric theories \cite{Bern:1994zx}. However due to the existence of the candidate Lorentz covariant 3-loop on-shell counterterm \cite{Kallosh:1980fi}, \cite{Howe:1981xy} it remains unclear why the UV divergence described by this candidate counterterm does not appear in the calculations of  \cite{Bern:2007hh}. If we find why this counterterm may be forbidden, it may help us to understand the UV properties of \N=8 SG and open a new way to address the possibility of its all-loop finiteness  \cite{Kallosh:2008mq}.

 One may try to give several different explanations why the 3-loop divergence is absent. For example, for  $L\geq 8$ loops the full non-linear  Lorentz covariant  counterterms are available \cite{Howe:1980th},  \cite{Kallosh:1980fi}. Meanwhile, only the linearized version of the 3-loop counterterm is known, so one could suspect that  the 3-loop finiteness discovered in \cite{Bern:2007hh} is related to the absence of the explicit non-linear 3-loop counterterm. However, we believe that this explanation is not valid since the finite 1-loop 4-point amplitudes are described by the same linearized superfields as  the 3-loop counterterm  \cite{Kallosh:2007ym}. Another explanation was suggested recently in \cite{Kallosh:2008mq}. It was noted there that the covariant $L$ loop counterterms may exist only if they can also be constructed by a different method,  using the CPT-conjugate light-cone superfield of \N=8 SG  \cite{Brink:1982pd}, \cite{Brink:2008qc}. As explained in  \cite{Kallosh:2008mq}, there are certain obstacles which may forbid  construction of counterterms by this method for any number of loops.  Recently it was suggested that one may overcome these difficulties and derive the 3-loop  counterterm in terms of the light-cone superfields   from the covariant counterterms \cite{KS}, but the actual construction is not yet available.   A significant progress in understanding \N=8 \, SG is required at this stage, in addition to direct computations of the higher-loop amplitudes.

The main purpose of this paper is to study the  bosonic 4-point amplitudes in \N=8 \, supergravity. We will present the explicit amplitudes with scalars which have not been computed so far.  This will allow us to verify the linearized \E \, symmetry of the 3-loop counterterm and 1-loop UV finite amplitudes.
We will translate the covariant 3-loop counterterm into a helicity formalism for 4-graviton, 4-vectors and 4-scalar amplitudes. We hope that this will help us eventually to test  the existence of the proper light-cone superfield 3-loop counterterm.

First, we will compute the 4-particle tree amplitudes for  vectors and scalars using the generating functional method proposed recently in \cite{Bianchi:2008pu}. The input into such functional is the 4-graviton amplitude, other amplitudes are given by the generating functional so that the supersymmetric Ward Identities  of \N=8 SG derived in
\cite{Bern:1998ug} are satisfied. We will also compute the 4-scalar tree amplitude using the old-fashioned method of Feynman graphs following from the \N=8 SG action. In this way we will have an independent confirmation of the generating function method of  \cite{Bianchi:2008pu}. From the explicit tree level scalar and vector 4-point amplitudes we will find the 3-loop UV divergent candidate 4-point amplitudes and the 1-loop UV finite scalar amplitudes, using the gravitational part of the counterterms and 1-loop amplitudes which are known explicitly.

A significant new input into the information about \N=8 SG was made in \cite{ArkaniHamed:2008gz} where the supersymmetric generalization of the theory to the complex momenta was proposed. It includes a shift in commuting as well as anticommuting spinors describing the external states. A new type of recursion relation between amplitudes was derived in \cite{ArkaniHamed:2008gz}  from  the vanishing of specific amplitudes at large complex momenta.
We will use an analogous  way of description of the counterterms, the candidate for the UV divergent amplitudes, and the 1-loop 4-point amplitudes to construct a supersymmetric generalization of  them for the case of complex momenta.

We are also interested in the 1-loop  4-point scalar amplitudes with regard to the \E \, symmetry. The related one-soft-scalar theorems for all tree amplitudes were established in \cite{Bianchi:2008pu} using Feynman graph methods, in \cite{ArkaniHamed:2008gz} using the complex deformation of momenta and the recursion relations and in \cite{KK} using the \E \,  Noether current constructed  in \cite{Kallosh:2008ic}. It is not known if this symmetry is also valid at the 1-loop level. After computing the  1-loop scalar amplitudes  we will be able to study their \E \, symmetry.
After we compute the explicit scalar amplitudes corresponding to  the 3-loop counterterm we will test its linearized \E \, symmetry. If the symmetry would be broken,  the counterterm would not be valid and this would be an explanation of the 3-loop finiteness. However, we will actually find that the 3-loop counterterm has a linearized \E \, symmetry. Also we will find that the 1-loop UV finite 4-point amplitudes are \E \, symmetric. Moreover, we  will study the  \E  symmetry of the 1-loop n-point amplitudes by looking at the 1-scalar-soft  limit of the 4-point function with complex momenta.

 The paper is organized as follows. In Sec. 2 we introduce the generating function for the 4-point amplitudes of \N=8 SG based on a perturbative expansion of the all positive helicity function ${\cal P}^L (1^+, 2^+, 3^+, 4^+)$.  We provide an input for this function at the tree level, $L=0$, at the 3-loop counterterm level, $L=3$ and at the UV finite 1-loop level, $L=1$. We also explain how to get all 4-point amplitudes starting from the function ${\cal P}^L (1^+, 2^+, 3^+, 4^+)$. In Sec. 3 we compute the 4-vector amplitude and the 2-scalar-2-vector  tree amplitudes. For the  4-scalar tree amplitudes we present in Sec. 4 the explicit expression derived from the generating function as well as from the Feynman rules. In Sec. 5 we use the tree level explicit amplitudes derived in the previous sections to find the helicity amplitudes corresponding to the 3-loop counterterm. In Sec. 6 the supersymmetric deformation of the generating function is studied. We find that the  3-loop counterterm does not depend on deformation and, in particular does not vanish at large complex momenta. The 1-loop amplitude does vanish at large $z$. In Sec. 7 we prove that both 1-loop finite 4-point amplitudes as well as 3-loop counterterms have an unbroken \E \, symmetry. Assuming that the \E \, Noether current is conserved for complex momenta we establish the relevant low-energy theorem for the $n$-point 1-loop amplitudes with generic $n$.
 In Sec. 8 we describe our findings and possible directions of the future work.


\section{The 4-point generating function in \N=8 \, SG}
Our main tool for calculating the amplitudes is the generating function method for \N=8 \,SG described in \cite{Bianchi:2008pu}\footnote{We  use notations from this paper. The spinor products are defined as follows: $ \langle ij \rangle = \tilde \lambda_{i \dot \alpha} \tilde \lambda_j ^{\dot \alpha} $ and $ [ ij ] =  \lambda_{i }^{ \alpha}  \lambda_{j  \alpha} $. }. We also use some features of the corresponding generating function in \N=4 \, YM as suggested in \cite{Drummond:2008vq}.

We suggest  to use the perturbative form of the all-loop generating function  for \N=8 \,SG 4-point function in the following form
\be
\Omega_4= \delta^4(\sum_i  \lambda_{\alpha i} \tilde \lambda_{\dot \alpha i}) \delta^{16} (\sum_i \tilde \lambda^{\dot \alpha}_{ i} \eta_{i A}) \; \sum _L \kappa^{2(L-1)}{\cal P}^L (1^+, 2^+, 3^+, 4^+) .
\label{Omega4} \ee
Here $i=1,2,3,4$ stands for 4 particles each with momentum $\lambda_{\alpha i } \tilde \lambda_{\dot \alpha i} =-(p_{\alpha \dot \alpha })_i$ and   $\eta_{iA}$ is a set of anticommuting variables and the index $A$ is an $SU(8)$ index, $A=1,...,8.$ The gravitational coupling constant is $\kappa^2$  and $L$ is the number of loops.
${\cal P}^L (1^+, 2^+, 3^+, 4^+)$  is the  function of spinors $\lambda_{\alpha i }$ and $ \tilde \lambda_{\dot \alpha i}$,  and it  has the following properties: it is symmetric in all 4 points and carries helicity $+2$ at each 4 points.

The generating function is manifestly supersymmetric under 16 supersymmetries and translation
\be
Q_A^{\dot \alpha} \; \Omega_4= Q^{A\alpha} \; \Omega_4 = P^{\dot \alpha \alpha} \; \Omega_4=0
\ee
which are defined as
\be
\tilde Q_A^{\dot \alpha}= \sum _i\tilde \lambda^{\dot \alpha}_{ i} \eta_{i A} \ , \qquad Q^{A\alpha}= \sum_i \lambda^\alpha_i {\partial \over \partial \eta_{iA}} \ , \qquad \{Q^{A\alpha}, \tilde Q_B^{\dot \alpha} \} = \delta^A{}_B \sum _i \tilde \lambda^{\dot \alpha}_{ i} \lambda^\alpha_i= \delta^A{}_B P^{\dot \alpha \alpha}
\ee
and  $ P^{\dot \alpha \alpha}\equiv  \sum _i P_i^{\dot \alpha \alpha}$.

The generating function is dimensionless. We  assign dimensions of $x$ to be -2 and of  $ \theta$ to be -1, this will translate into dimension of  $\delta^4(  \lambda \tilde \lambda) \sim \int d^4x e^{i\lambda \tilde \lambda x} $ to be -8 and that of $\delta^{16} ( \tilde \lambda \eta)\sim \int d^{16 }\theta e^{\theta \tilde \lambda \eta}$ to be +16.
Dimension of $\kappa$ is -2. At every loop order $\kappa^{2(L-1)}{\cal P}^L (1^+, 2^+, 3^+, 4^+)$ has dimension -8 and ${\cal P}^L (1^+, 2^+, 3^+, 4^+)$ has dimension $4(L-3)$.

For the tree amplitudes we may use the following form
\be
{\cal P}^{tree} (1^+, 2^+, 3^+, 4^+)=\kappa^{-2}{\cal P}^0 (1^+, 2^+, 3^+, 4^+)=
{s_{12} s_{14}\over \kappa^2 s_{13}} {1\over
 \left (\langle 1 2\rangle  \langle 23\rangle  \langle 34\rangle  \langle 41\rangle \right  )^{2}} .
\ee
Here $-s_{ij}= (p_i+p_{j})^2=  \langle ij\rangle [ i j]$. The product of spinor angle brackets provides helicity $+2$ at each 4 points.  In terms of  the standard Mandelstam variables $s_{12}=s, s_{13}=t, s_{14}=u$ it is
 \be
{\cal P}^{tree} (1^+, 2^+, 3^+, 4^+)= {s \; u \over \kappa^2  t} {1\over
 (\langle 1 2\rangle  \langle 23\rangle  \langle 34\rangle  \langle 41\rangle )^{2}} .
\ee
The dependence on Mandelstam variables, $s, t, u$ in the form ${s \; u\over \kappa^2 t}$ is present in \N=8 \, supergravity as different from  \N=4 \, YM where the analogous ${\cal P}$ function  is
$${\cal P} _{YM}^{tree} (1^+, 2^+, 3^+, 4^+)= {1\over
 \langle 1 2\rangle  \langle 23\rangle  \langle 34\rangle  \langle 41\rangle }
 $$
 and carries helicity +1 at each point.

For the 3-loop counterterm appropriate for the helicity formalism as suggested in \cite{Kallosh:2007ym,Kallosh:2008mq}  we take
 \be
{\cal P}^{3-loop}_{UV} (1^+, 2^+, 3^+, 4^+)= \kappa^4 {\cal P}^{3}_{UV}  = \left  (  {\kappa^2 s_{12} s_{14} \over
 \langle 1 2\rangle  \langle 23\rangle  \langle 34\rangle  \langle 41\rangle }\right)^2
\ee
or
\be
{\cal P}^{3-loop}_{UV} (1^+, 2^+, 3^+, 4^+)= \kappa^6 s t u \, {\cal P}^{tree} (1^+, 2^+, 3^+, 4^+) .
\ee
For the 1-loop UV finite amplitude \cite{Green:1982sw,Bern:1998ug} we take according to \cite{Kallosh:2007ym},  \cite{Kallosh:2008mq}
 \be
{\cal P}^{1-loop} (1^+, 2^+, 3^+, 4^+)= {\cal P}^{3}_{UV} F^{box}
\ee
where ${\cal P}^{3}_{UV}$ is dimensionless
\be
{\cal P}^{3}_{UV}  (1^+, 2^+, 3^+, 4^+)= \left  (  {s_{12} s_{14} \over
 \langle 1 2\rangle  \langle 23\rangle  \langle 34\rangle  \langle 41\rangle }\right)^2.
\ee
The box integral is represented by the functions ${\cal I}_4^{1-loop}$  \cite{Green:1982sw,Bern:1998ug}  of dimension -8 :
\be
F^{box}= \left ({\cal I}_4^{1-loop} (s_{12}, s_{23}) +{\cal I}_4^{1-loop} (s_{12}, s_{13})+{\cal I}_4^{1-loop} (s_{23}, s_{13})\right) .
\label{Fbox}\ee
The explicit form of these integrals can be taken from eq. (4.25) of \cite{Bern:1993kr} or from Appendix B of \cite{Bern:2005iz}.
The IR divergence is taken care by using dimension $d=4-2\epsilon$ and computing the ${1\over \epsilon}$ terms as well as $\epsilon$-independent terms. The terms ${1\over \epsilon^2}$ cancel between the 3 terms above. The individual entry is
\be
{\cal I}_4^{1-loop} (s, t)= -{1\over (-s)^{1+\epsilon} t}\left [{4\over \epsilon^2} + {2 \ln (s/t)\over \epsilon} -4\pi^2 /3 \right] .
\label{I4}\ee

\subsection{The amplitudes}
 To extract the 4-point amplitudes from the generating function one has to pull out  16
 anticommuting variables \cite{Bianchi:2008pu}: each external state is  associated with  a differential operator. We present them in the following order:  positive helicity graviton, $   b_+(i)$, positive helicity gravitino, $f_+^A(i)$, etc. all the way till  negative helicity  graviton, $b^-(i)$, representing all physical states of the CPT conjugate supermultiplet of \N=8 \, SG:
 \begin{align}
&b_+(i)\leftrightarrow 1,\ f_+^A(i)\leftrightarrow \frac{\partial}{\partial\eta_{iA}},\ b_+^{AB}(i)\leftrightarrow\frac{\partial^2}{\partial\eta_{iA}\partial\eta_{iB}},\ f_+^{ABC}(i)\leftrightarrow\frac{\partial^3}{\partial\eta_{iA}\partial\eta_{iB}\partial\eta_{iC}},\cr
&\ b^{ABCD}(i)\leftrightarrow\frac{\partial^4}{\partial\eta_{iA}\partial\eta_{iB}\partial\eta_{iC}\partial\eta_{iD}},\ f_{ABC}^-(i)\leftrightarrow-\frac{1}{5!}\epsilon_{ABCDEFGH}\frac{\partial^5}{\partial\eta_{iD}...\partial\eta_{iH }},\cr
&b_{AB}^-(i)\leftrightarrow\frac{1}{6!}\epsilon_{ABCDEFGH}\frac{\partial^6 }{\partial\eta_{iC}...\partial\eta_{iH}},\ f_A^-(i)\leftrightarrow-\frac{1}{7!}\epsilon_{ABCDEFGH}\frac{\partial^7}{\partial\eta_{iB}...\partial\eta_{iH}},\cr
&b^-(i)\leftrightarrow\frac{1}{8!}\epsilon_{ABCDEFGH}\frac{\partial^8}{\partial\eta_{iA}...\partial\eta_{iH}}.
\end{align}
One then acts with these operators on the generating function to get the amplitude.
We may  present the generating function in the form given in \cite{Bianchi:2008pu}
\be
\Omega_4=\frac{\hat M_4(1^-,2^-,3^+,4^+)}{\langle12\rangle^8 }\prod_{A=1}^8\sum_{i>j\ge1}^4\langle ij\rangle\eta_{iA}\eta_{jA}
\label{Omega}\ee
  where
\be
\hat M(1^-, 2^-, 3^+, 4^+)= \delta^4(\sum_i  \lambda_{\alpha i} \tilde \lambda_{\dot \alpha i})  {\cal P} (1^+, 2^+, 3^+, 4^+) \langle12\rangle^8
\label{BEF} \ee
is dimensionless. Here
\be
\hat M(1^-, 2^-, 3^+, 4^+)= \delta^4(\sum_i  \lambda_{\alpha i} \tilde \lambda_{\dot \alpha i}) M(1^-, 2^-, 3^+, 4^+)
\ee
 and $M(1^-, 2^-, 3^+, 4^+)$ is a physical non-vanishing amplitude with two negative and two positive helicity gravitons \footnote{This explains why in  \cite{Bianchi:2008pu} $\Omega_4$  was given in the form (\ref{Omega}).
When  extracting  8 powers of $\eta_{1A}$ and 8 powers of $\eta_{2B}$ to get two gravitons of negative helicity, one gets $  \langle12\rangle^8 $.
 } .

In the tree approximation the 4-graviton amplitude in helicity formalism is
\be
M^{tree}(1^-,2^-,3^+,4^+)=\frac{\langle12\rangle^4[34]^4 }{\kappa^2 stu}.
\ee
The 4-graviton amplitude which for a long time was considered as  a candidate for the 3-loop divergence \cite{Kallosh:1980fi}, \cite{Howe:1981xy},  \cite{Kallosh:2008mq}  is given by the square of the Bel-Robinson tensor, $R_{\alpha \beta\gamma\delta}  \bar R_{\dot \alpha \dot \beta\dot \gamma\dot \delta} R^{\alpha \beta\gamma\delta}  \bar R^{\dot \alpha \dot \beta\dot \gamma\dot \delta}$
which in helicity formalism is \cite{Kallosh:2008mq}
\be
M_{UV}^{3-loop}(1^-,2^-,3^+,4^+)=\kappa^4 \langle12\rangle^4[34]^4 .
\ee
For the 1-loop UV finite amplitude we get
\be
M^{1-loop}(1^-,2^-,3^+,4^+)=M_{UV}^{3-loop}(1^-,2^-,3^+,4^+){1\over \kappa^4} F^{box}=
\langle12\rangle^4[34]^4    F^{box}\ ,
\ee
where $F^{box}$ is defined in eq. (\ref{Fbox}).

The supersymmetric partners of the 4-graviton 3-loop amplitude can be inferred either from the local superaction integral \cite{Kallosh:1980fi}, \cite{Howe:1981xy} by performing the integration over the anti-commuting variables or, alternatively, by using the generating functional method \cite{Bianchi:2008pu}.
The 1-loop amplitudes including  scalars will allow us to study their \E \, symmetry.


\section{Amplitudes with vectors}
In this section we discuss only tree amplitudes and use notation where $\kappa^2=1$.

\subsection{4-vector and 2-scalar-2-vector tree amplitude}

Our first example how to use the generating function to get the 4-point vector amplitude is given by the 4-vector amplitude
\be
\langle b_{AB}^-b_{CD}^-b_+^{EF}b_+^{GH}\rangle= \frac{1}{6!}\epsilon_{ABA'B'C'D'E'F'}\frac{\partial^6}{\partial\eta_{1A'}...\partial_{1F'}}\frac{1}{6!}\epsilon_{CDKLMNPQ}\frac{\partial^6}{\partial\eta_{2K}...\partial_{2Q}} \frac{\partial^2}{\partial\eta_{3E}\partial_{3F}}\frac{\partial^2}{\partial\eta_{4G}\partial_{4H}}\Omega_4,
\ee
Using Young tableaux the tensor product $\overline{28_A}\otimes \overline{28_A} \otimes 28_A\otimes 28_A$ can be shown to contain three singlets. If we manage to find three linearly independent invariant tensors that transforms in the representation above we can use these in an ansatz for the four point function. What is then the expressions for these singlets? Singlets in SU(8) must be constructed using $\delta^A_B$, $\epsilon_{ABCDEFGH}$ and $\epsilon^{ABCDEFGH}$. For a given index structure we then pick matching $\delta$'s and $\epsilon$'s, write down indices and finally (anti)symmetrize if necessary. In our case a good ansatz is
\be
\langle b_{AB}^-b_{CD}^-b_+^{EF}b_+^{GH}\rangle=a\delta_{AB}^{EF}\delta_{CD}^{GH}+b\delta_{AB}^{GH}\delta_{CD}^{EF}+c\delta_{ABCD}^{EFGH},
\ee
where $\delta_{AB}^{EF}$ and $\delta_{ABCD}^{EFGH}$ are antisymmetrized delta functions that take the values 0,1 and -1. To fix the values of $a$, $b$ and $c$ we calculate, using the generating functional method, the amplitudes where $\{A...H\}$ takes the values $\{12131213\}$, $\{12131312\}$, and $\{12342341\}$ respectively. Since the terms can be made to vanish independently we know that they aren't linear combinations of each other. This guarantees us that the ansatz above is sufficient. For the first set of indices all terms except the first vanishes and we get for the tree amplitude
\be
a=\frac{[34]^4}{stu\langle12\rangle^4 }\langle12\rangle^5\langle14\rangle\langle23\rangle\langle34\rangle=
-\frac{\langle12\rangle^2[34]^2 }{t}.
\ee
Similar calculations the other sets of indices give
\be
b=-\frac{\langle12\rangle^2[34]^2 }{u}\text{ and } c=-\frac{\langle12\rangle^2[34]^2 }{s}
\ee
and thus the 4-point vector amplitude at the tree level is
\be
\langle b_{AB}^-b_{CD}^-b_+^{EF}b_+^{GH}\rangle=-\langle12\rangle^2[34]^2\left[\frac{1}{t}\delta_{AB}^{EF}\delta_{CD}^{GH}+\frac{1}{u}\delta_{AB}^{GH}\delta_{CD}^{EF}+\frac{1}{s}\delta_{ABCD}^{EFGH}\right].
\label{eqn:4vectors}
\ee


\subsection{2-vector-2-scalar tree amplitude}
Here we have to compute
\begin{align}
\langle b_{AB}^-b_+^{CD}b^{EFGH}b^{IJKL} \rangle=\frac{1}{6!}\epsilon_{ABMNPQRS}\frac{\partial^6}{\partial\eta_{1M}...\partial_{1S}}\frac{\partial^2}{\partial\eta_{2C}\partial_{2D}}\frac{\partial^4}{\partial\eta_{3E}\partial_{3F}\partial_{3G}\partial_{3H} }\frac{\partial^4}{\partial\eta_{4I}\partial_{4J}\partial_{4K}\partial_{4L} }\Omega_4.
\end{align}
In this case we know that the four point function is written in terms of the singlets in $\overline{28_A}\otimes28_A\otimes70_A\otimes70_A$. Using Young tableaux we see that the tensor product contains three singlets and we thus make the ansatz
\be
\langle b_{AB}^-b_+^{CD}b^{EFGH}b^{IJKL} \rangle =a\delta^{CD}_{AB}\epsilon^{EFGHIJKL}+b\delta_{AB}^{[EF }\epsilon^{GH]IJKLCD}+c\delta_{AB}^{[IJ}\epsilon^{KL]EFGHCD}
\ee
where antisymmetrization is defined with a factor $1/4!$. Again, the values of $a$, $b$ and $c$ can be fixed by choosing $\{A...L\}$ in an appropriate way, in this case $\{161612345678\}$, $\{121312452678\}$ and $\{121324561278\}$ respectively. We fix $a$  by evaluating the first combination:
\begin{align}
a&=\langle b_{16}^-b_+^{16}b^{1234}b^{5678} \rangle=\frac{[34]^4}{stu\langle12\rangle^4 }\langle13\rangle^3\langle14\rangle^3\langle23\rangle\langle24\rangle=-\frac{\langle13\rangle^2[23]^2 }{s}.
\end{align}
Similar calculations gives us
\be
b=-3!\frac{\langle13\rangle^2[23]^2 }{t} \quad  \rm { and } \quad c=-3!\frac{\langle13\rangle^2[23]^2}{u}.
\ee
Thus the result for the 2-vector-2scalar tree amplitude is:
\be \langle b_{AB}^-b_+^{CD}b^{EFGH}b^{IJKL} \rangle = -\langle13\rangle^2 [23]^2 \left[ \frac{1}{s}\delta^{CD}_{AB}\epsilon^{EFGHIJKL} + \frac{3!}{t}\delta_{AB}^{[EF }\epsilon^{GH]IJKLCD}+\frac{3!}{u}\delta_{AB}^{[IJ}\epsilon^{KL]EFGHCD}\right] \label{2+2tree} .\ee


\section{4-Scalar tree amplitude}

In this section again we discuss only tree amplitudes and therefore use notation where $\kappa^2=1$.
The 4-scalar amplitude we will compute using two methods. The first one is the same as we used for the 4-vector and 2-vector-2scalar case above, namely we will use the generating function. The second computation of the 4-scalar tree amplitude we will perform using the Feynman rules of \N=8 \, SG. The reason for doing this double computation is the following. In \N=4 \, YM theory there were many computations using the standard Feynman diagrams for gauge fields with polarization operators and the results have been shown to lead to helicity amplitudes in the form which is most useful for \N=4 \, YM theory. In \N=8 \, SG no such computations of the 4-scalar tree level amplitude have been performed neither in the helicity formalism using the generating function nor using the standard Feynman rules.

Since we will use the explicit form of the 4-scalar amplitude at the tree level to identify the 4-scalar candidate UV divergent amplitude at the 3-loop level, we would like to make sure that our expression passes the test: it is the same when computed via the generating function method as the one computed from the Feynman rules.

\subsection{Generating functional}
The expression for the 4-scalar amplitude is
\bea
&&\langle b^{ABCD}b^{EFGH}b^{IJKL}b^{MNPQ}\rangle= \frac{\partial^4}{\partial\eta_{1A}\partial\eta_{1B}\partial\eta_{1C}\partial\eta_{1D}} \frac{\partial^4}{\partial\eta_{2E}\partial\eta_{2F}\partial\eta_{2G}\partial\eta_{2H}}\nonumber\\
\nonumber\\
&& \frac{\partial^4}{\partial\eta_{3I}\partial\eta_{3J}\partial\eta_{3K}\partial\eta_{3L}} \frac{\partial^4}{\partial\eta_{4M}\partial\eta_{4N}\partial\eta_{4P}\partial\eta_{4Q}} \Omega_4 .
\eea
The tensor product $70_A\otimes70_A\otimes70_A\otimes70_A$ contains five singlet irreducible representations. The six singlets
\begin{align}
O_1^{A...Q}=\epsilon^{ABCDEFGH}\epsilon^{IJKLMNPQ},\ &O_4^{A...Q}=\frac{1}{4!^4}\sum_{\textit{perm} }(-1)^\textit{perm}\epsilon^{1_11_21_31_43_13_24_34_4} \epsilon^{2_12_22_32_44_14_23_33_4}\cr
O_2^{A...Q}=\epsilon^{ABCDIJKL}\epsilon^{EFGHMNPQ},\ &O_5^{A...Q}=\frac{1}{4!^4}\sum_{\textit{perm} }(-1)^\textit{perm}\epsilon^{1_11_21_31_42_12_24_34_4}\epsilon ^{3_13_23_33_44_14_22_32_4}\cr
O_3^{A...Q}=\epsilon^{ABCDMNPQ}\epsilon^{EFGHIJKL},\ &O_6^{A...Q}=\frac{1}{4!^4}\sum_{\textit{perm} }(-1)^\textit{perm}\epsilon^{1_11_21_31_42_12_24_34_4} \epsilon^{4_14_24_34_43_13_22_32_4},
\end{align}
where $\{1_1,1_2,1_3,1_4\}$ are permutations of $\{A,B,C,D\}$ and so forth, are related trough
\be
O_1+O_2+O_3=12\left(O_4+O_5+O_6\right).
\label{eqn:operatorrelation}
\ee
This means that the ansatz
\be
\langle b^{ABCD}b^{EFGH}b^{IJKL}b^{MNPQ}\rangle=aO_1^{A...Q} + bO_2^{A...Q} + cO_3^{A...Q} + dO_4^{A...Q} + eO_5^{A...Q} + fO_6^{A...Q}
\ee
contains five linearly independent singlets and thus should capture the four point function. To fix the coefficients we calculate the four point function when $\{A...Q\}$ takes the six values \{1234123456785678\}, \{1234567812345678\}, \{1234567856781234\}, \{1234127856345678\}, \{123\-4127856785634\} and \{1234567812785634\}, giving us
\begin{align}
b+c+\frac{d}{3!}&=\frac{[34]^4}{\langle12\rangle^4stu }\langle12\rangle^4\langle34\rangle^4=\frac{s^3}{tu}\cr
a+c+\frac{e}{3!}&=\frac{[34]^4}{\langle12\rangle^4stu }\langle13\rangle^4\langle24\rangle^4=\frac{t^3}{su}\cr
a+b+\frac{f}{3!}&=\frac{[34]^4}{\langle12\rangle^4stu }\langle14\rangle^4\langle23\rangle^4=\frac{u^3}{st} \cr
c+\frac{1}{3!^2}\left(d+e+f\right)&=\frac{[34]^4 }{\langle12\rangle^4stu}\langle12\rangle^2\langle13\rangle^2\langle24\rangle^2\langle34\rangle^2=\frac{st}{u}\cr
b+\frac{1}{3!^2}\left(d+e+f\right)&=\frac{[34]^4 }{\langle12\rangle^4stu}\langle12\rangle^2\langle14\rangle^2\langle23\rangle^2\langle34\rangle^2=\frac{su}{t}\cr
a+\frac{1}{3!^2}\left(d+e+f\right)&=\frac{[34]^4 }{\langle12\rangle^4stu}\langle13\rangle^2\langle14\rangle^2\langle23\rangle^2\langle24\rangle^2=\frac{tu}{s}.
\end{align}
Because of (\ref{eqn:operatorrelation}) this system is degenerate. One particulary symmetric solution is
\be
a=\frac{tu}{s},\ b=\frac{su}{t},\ c=\frac{st}{u},\ d=2(3!)s,\ e=2(3!)t ,\ f=2(3!)u
\ee
giving us the correlation function
\begin{align}
\langle b^{ABCD}&b^{EFGH}b^{IJKL}b^{MNPQ}\rangle=\cr
&\frac{tu}{s}\epsilon^{ABCDEFGH}\epsilon^{IJKLMNPQ}+\frac{su}{t}\epsilon^{ABCDIJKL}\epsilon^{EFGHMNPQ}
+\frac{st}{u}\epsilon^{ABCDMNPQ}\epsilon^{EFGHIJKL}\cr
&+\frac{1}{2(4!)^3}\sum_{\textit{perm}}(-1)^\textit{perm}\left[s\epsilon^{1_11_21_31_43_13_24_34_4}\epsilon^{2_12_22_32_44_14_23_33_4}+t\epsilon^{1_11_21_31_42_12_24_34_4}\epsilon^{3_13_23_33_44_14_22_32_4}+\right.\cr
&\left.u \epsilon^{1_11_21_31_42_12_23_33_4}\epsilon^{4_14_24_34_43_13_22_32_4}\right].
\label{eqn:4scalars}
\end{align}


\subsection{\N=8 SG tree Feynman graphs for the 4-scalar amplitude}

There are two types of Feynman graphs which form a tree level 4-scalar amplitude. There is a cubic scalar-scalar-graviton interaction with the graviton exchange and a contact 4-scalar interaction. To establish the relevant Feynman rules we need the gravity part and  the scalar part of the action \cite{Cremmer:1979up,Kallosh:2008ic}
\be
\mathcal{L}_{R+sc}=-{1\over 2} e R -\frac{e}{96}\mathcal{A}^{ABCD}_{\mu}\mathcal{A}^\mu_{ABCD}= -{1\over 2} e R -\frac{e}{12}\textup{Tr}\left(\frac{1}{1-y\bar y }\partial_\mu y\frac{1}{1-\bar y y}\partial^\mu\bar y\right),
\ee
where we use the convention $\kappa^2=1$ and
\begin{align}
y_{AB,CD}= \phi_{ABEF}\left(\frac{\textup{tanh}\sqrt{\bar \phi\phi/8}}{\sqrt{\bar \phi\phi}}
\right)^{EF}_{CD}=\frac{1}{\sqrt{8}}\left(\phi_{ABCD}-\frac{1}{4!}\phi_{ABEF}\bar\phi^{EFGH}\phi_{GHCD}\right)+\mathcal{O}(\phi^5)
\end{align}
and
\begin{equation}
\phi_{ABCD}=\frac{1}{4!}\epsilon_{ABCDEFGH}\bar\phi^{EFGH}.
\label{eqn:phiduality}
\end{equation}
Using this we can expand
\begin{align}
&-\frac{e}{12}\textup{Tr}\left(\frac{1}{1-y\bar y }\partial_\mu y\frac{1}{1-\bar y y}\partial^\mu\bar y\right)=-\frac{e}{12} \textup{Tr}\left(\left(1+y\bar y\right)\partial_\mu y\left(1+\bar y y\right)\partial^\mu \bar y+\mathcal{O}(y^6)\right)\cr
&=-\frac{e}{4(4!)}\textup{Tr}\left(\partial_\mu\phi\partial^\mu\bar\phi+\frac{1}{8}\phi\bar\phi\partial_\mu\phi\partial^\mu\bar \phi+\frac{1}{8}\partial_\mu\phi\bar\phi\phi\partial^\mu\bar\phi-\frac{1}{4!}\partial_\mu(\phi\bar \phi\phi)\partial^\mu\bar\phi-\frac{1}{4!}\partial^\mu\phi\partial^\mu(\bar \phi\phi\bar\phi)+\mathcal{O}(\phi^6)\right)\cr
&=-\frac{e}{4(4!)}\textup{Tr}\left(\partial_\mu\phi\partial^\mu\bar\phi+\frac{1}{4!}\left(\partial_\mu\phi\partial^\mu\bar\phi\phi\bar\phi-\partial_\mu\phi\bar\phi\partial^\mu\phi\bar\phi+\partial_\mu\phi\bar\phi\phi\partial^\mu\bar\phi -\phi\partial_\mu\bar\phi\phi\partial^\mu\bar\phi+\mathcal{O}(\phi^6)\right)\right).
\label{eqn:y-expansion}
\end{align}
The first term will give a kinetic term and a graviton interaction term whereas the other give scalar interactions. Let us first focus on the graviton exchange. Using $\sqrt{-g}=e$ and (\ref{eqn:phiduality}) we get
\be
\mathcal{L}_{sc,g}\simeq-\frac{\sqrt{-g}g^{\mu\nu} }{4}\frac{1}{4!^2}\epsilon^{ABCDEFGH}\partial_\mu\phi_{ABCD}\partial_\nu\phi_{EFGH},
\ee
where we have written the metric dependence explicitly. To calculate the contribution coming from the graviton exchange we  use $g_{\mu\nu}=\eta_{\mu\nu}+h_{\mu\nu}$, $g^{\mu\nu}=\eta^{\mu\nu}-h^{\mu\nu}+...$ and
\be
\mathcal{L}_{sc,g}\simeq\frac{1}{4}\frac{1}{4!^2}\epsilon^{ABCDEFGH}\left(-\partial_\mu\phi_{ABCD}
\partial^\mu\phi_{EFGH}+
(h^{\mu\nu}- \eta^{\mu\nu}\frac{1}{2}h^\lambda_\lambda)  \partial_\mu\phi_{ABCD}\partial_\nu\phi_{EFGH} \right).
\ee
We see that for the kinetic term to be appropriately normalized we have to do the rescaling $\phi\rightarrow \sqrt{2}\phi$ (One way to see this is by recombining all equivalent terms using (\ref{eqn:phiduality}). This gives a theory of unconstrained complex fields with a kinetic term who's coefficient is 1/2). The Feynman rules for the graviton propagator in the Feynman gauge and the scalar-scalar-graviton vertex are
\begin{align}
&
\parbox{20mm}{\includegraphics{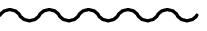}} \qquad \qquad =\qquad  \frac{i\left(\eta_{\mu\alpha}\eta_{\nu\beta}+\eta_{\mu\beta }\eta_{\nu\alpha}-\eta_{\mu\nu }\eta_{\alpha\beta}\right) }{2 \, k^2  }
\cr
\cr
&
\parbox{20mm}{\includegraphics{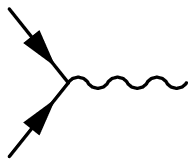}}\qquad \qquad =\qquad  -i\left(\eta_{\mu\nu}p_1\cdot p_2-p_{1\mu }p_{2\nu}-p_{1\nu }p_{2\mu }\right)\epsilon^{ABCDEFGH}.\cr
\end{align}

 We can now evaluate the diagram

\begin{align}
&
  \parbox{20mm}{
	\includegraphics{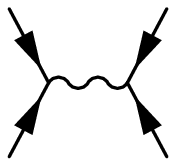}}\qquad =\qquad   (-i)\left(\eta_{\mu\nu}p_1\cdot p_2-p_{1\mu}p_{2\nu }-p_{1\nu }p_{2\mu }\right)\epsilon^{ABCDEFGH}\times\cr\cr
&\times (-i)\left(\eta_{\alpha\beta }p_3\cdot p_4-p_{3\alpha }p_{4\beta }-p_{3\beta }p_{4\alpha }\right)\epsilon^{IJKLMNPQ}i\frac{\eta^{\mu\alpha}\eta^{\nu\beta}+\eta^{\mu\beta }\eta^{\nu\alpha}-\eta^{\mu\nu }\eta^{\alpha\beta}}{2\, (p_1+p_2)^2}=\cr
&= {i}\frac{tu}{s}\epsilon^{ABCDEFGH}\epsilon^{IJKLMNPQ}.
\end{align}
Including the other two channels gives the amplitude
\begin{align}
\langle b^{ABCD}&b^{EFGH}b^{IJKL}b^{MNPQ}\rangle_{graviton}
=\frac{tu}{s}\epsilon^{ABCDEFGH}\epsilon^{IJKLMNPQ}\cr &+\frac{su}{t}\epsilon^{ABCDIJKL}\epsilon^{EFGHMNPQ}
+\frac{st}{u}\epsilon^{ABCDMNPQ}\epsilon^{EFGHIJKL}
\end{align}
which exactly corresponds to the second line of (\ref{eqn:4scalars}). To get the rest of the four point function we have to turn to the last four terms in (\ref{eqn:y-expansion}). First note that
\begin{align}
\textup{Tr}\left(\phi\bar\phi\phi\bar\phi\right)&=\frac{1}{4!^2}\phi_{ABCD}\epsilon^{CDEFGHIJ}\phi_{GHIJ}\phi_{EFKL}\epsilon^{KLABMNPQ}\phi_{MNPQ}\cr
&=\frac{1}{4!^2} \epsilon^{1_31_42_12_22_32_43_13_2}\epsilon^{3_33_44_14_24_34_41_11_2}\phi_{1_11_21_31_4}\phi_{2_12_22_32_4}\phi_{3_13_23_33_4}\phi_{4_14_24_34_4}\cr
&=\frac{1}{4!^2} \epsilon^{1_11_21_31_42_12_24_34_4}\epsilon^{3_13_23_33_44_14_22_32_4}\phi_{1_11_21_31_4}\phi_{2_12_22_32_4}\phi_{3_13_23_33_4}\phi_{4_14_24_34_4},
\end{align}
which has the same index structure as the second term in the third line of (\ref{eqn:4scalars}).  We now want to contract (\ref{eqn:y-expansion}) with the external fields. We see that when an external field is contracted with $\bar\phi$ it give rise to a term with all four of its indices in the same $\epsilon$. Contracting the first external scalar to the first field in $\phi\bar\phi\phi\bar\phi$, the second to the second etc. give, adding a factor of $-i$ and keeping the rescaling in mind,
\begin{align}
\frac{i}{(4!)^4}\sum_{perm}&\epsilon^{1_11_21_31_42_12_24_34_4}\epsilon^{3_13_23_33_44_14_22_32_4}\times\cr\times
&\left((-ip_{1\mu })(-ip_{2}^\mu)-(-ip_{1\mu})(-ip_3^\mu)+(-ip_{1\mu})(-ip_4^\mu))-(-ip_{2\mu})(-ip_4^\mu)\right)\cr
=\frac{i}{16(4!)^3}\sum_{perm}&\epsilon^{1_11_21_31_42_12_24_34_4}\epsilon^{3_13_23_33_44_14_22_32_4}t.
\end{align}
Adding up all 4! contractions gives us the amplitude

\begin{align}
\parbox{20mm}{
\includegraphics{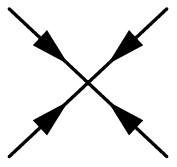}
}=\frac{1}{2(4!)^3}\sum_{perm}&\left(\epsilon^{1_11_21_31_43_13_24_34_4}\right.\epsilon^{2_12_22_32_44_14_23_33_4}s\cr
&+\epsilon^{1_11_21_31_42_12_24_34_4}\epsilon^{3_13_23_33_44_14_22_32_4}t\cr
&\left.+\epsilon^{1_11_21_31_42_12_23_33_4}\epsilon^{4_14_24_34_43_13_22_32_4}u\right)
\end{align}
which matches the last two lines in (\ref{eqn:4scalars}).

\section{3-loop candidate divergences of the  4-point amplitudes}
Starting from the 4-graviton candidate for the 3-loop divergence and using the relation between the tree 4-amplitudes and 3-loop counterterms, we are ready to present here the 4-vector, 2-vector-2-scalar and 4-scalar supersymmetric partners of the 4-graviton candidate for the 3-loop divergence. The 4-graviton amplitude part is well known
\be
M_{4grav}^{3-loop}(1^-,2^-,3^+,4^+)_{UV}=\kappa^4 \langle12\rangle^4[34]^4  .
\ee
The 4-vector part is
\be
M_{4vec}^{3-loop}( b_{AB}^-, b_{CD}^-, b_+^{EF}, b_+^{GH})_{UV} =-\kappa^4 \langle12\rangle^2[34]^2\left[s\, u\, \delta_{AB}^{EF}
\delta_{CD}^{GH}+s\, t \, \delta_{AB}^{GH}\delta_{CD}^{EF}+t\, u\, \delta_{ABCD}^{EFGH}\right].
\label{4vectors}
\ee
The amplitude for 2 vectors and 2 scalars is
\bea
&&M_{2vec2sc}^{3-loop}( b_{AB}^-, b_+^{CD}, b^{EFGH}, b^{IJKL} )_{UV} = -\kappa^4 \langle13\rangle^2 [23]^2 \times \nonumber\\
\nonumber\\
&&\left[ t\, u\, \delta^{CD}_{AB}\epsilon^{EFGHIJKL} + 3! \, s\, u\, \delta_{AB}^{[EF }\epsilon^{GH]IJKLCD}+3! \, s\, t\, \delta_{AB}^{[IJ}\epsilon^{KL]EFGHCD}\right]  .\label{2+2}\eea
Finally, the 4 scalar amplitude is
\begin{align}
&M_{4sc}^{3-loop} ( b^{ABCD},  b^{EFGH}, b^{IJKL}, b^{MNPQ}\rangle_{UV}=\cr
&\kappa^4(t^2\, u^2\, \epsilon^{ABCDEFGH}\epsilon^{IJKLMNPQ}+s^2\, u^2\, \epsilon^{ABCDIJKL}\epsilon^{EFGHMNPQ}
+s^2 \, t^2\, \epsilon^{ABCDMNPQ}\epsilon^{EFGHIJKL})\cr
&+\frac{\kappa^4}{2(4!)^3}\sum_{\textit{perm}}(-1)^\textit{perm}\left[s^2\, t\, u\, \epsilon^{1_11_21_31_43_13_24_34_4}\epsilon^{2_12_22_32_44_14_23_33_4}+ s\, t^2\, u\, \epsilon^{1_11_21_31_42_12_24_34_4}\epsilon^{3_13_23_33_44_14_22_32_4}+\right.\cr
&\left.s\, t\, u^2\,  \epsilon^{1_11_21_31_42_12_23_33_4}\epsilon^{4_14_24_34_43_13_22_32_4}\right].
\label{4scalars}
\end{align}

Various bosonic  amplitudes above presents a UV divergent part of the  manifestly supersymmetric invariant depending on the scalar superfield
$W_{ i_1...i_4}(x, \theta)$ whose first component is the scalar field $\phi_{ i_1...i_4}(x)$ \cite{Kallosh:1980fi}, \cite{Howe:1981xy}:
\bea
S^{3-loop}_{UV}=A_3 {\kappa^{4}\over \epsilon} \int d^4 x \, D^{[i_1...i_4][j_1...j_4]} \bar D^{[k_1...k_4][l_1...l_4]}
 \times (W_{ i_1...i_4}  W_{ j_1...j_4}  W_{ k_1...k_4}  W_{ l_1...l_4} ).
\label{superaction}\eea
The integration measure is an $SU(8)$ tensor and it corresponds to the integration over the half of the 32 superspace anti-commuting coordinates. The kernel is an $SU(8)$ tensor corresponding to a square Young tableaux,
which is the $\underline{232848}$ representation of the $SU(8)$.
This is a structure of the UV divergence,  with $\epsilon= d-4$,  at the 3-loop order in \N=8 supergravity.

If one would perform explicitly the 16 Grassmann integration in (\ref{superaction}), one would find all 4-point bosonic amplitudes which we have already obtained using the generating function or Feynman rules and the known relation between
the tree diagrams and the 3-loop counterterm in the 4-graviton amplitude.

\section{Supersymmetric complex deformation of the generating function}

A method of the complex-valued
shift  of the momenta of a pair of external particles was proposed in \cite{Britto:2005fq}.
During the last few years it was used for the on-shell amplitude computations in gauge theories \cite{Bern:2005cq}. More specifically the BCFW shift   is performed on two commuting spinors, specifying the external momenta, for example, on $\lambda_\alpha(p_1)$ and on $\tilde \lambda_{\dot \alpha}(p_2)$.

A supersymmetric extension of the BCFW  complex-valued shift of momenta was proposed in \cite{ArkaniHamed:2008gz}. Each external state is characterized by the momentum $p_{i \, \alpha \dot \alpha} =- \lambda_\alpha(p_i) \tilde \lambda_{\dot \alpha}(p_i)$ as well as by an  anti-commuting spinor $\eta_{i A}$. Consider the following shift of particular 2 points, $k$ and $l$ which could be any of the 4 points in the amplitude.
\be
\lambda_k(z)= \lambda_k +z \lambda_l \ ,  \qquad \tilde \lambda_l (z) = \tilde \lambda_l -z \tilde \lambda_k  \qquad \eta_{k} (z) = \eta_{k} +z \eta_{l}.
\ee
Here we are using a short notation $\lambda_\alpha (p_i)\equiv \lambda_i$ and
$\tilde \lambda_{\dot \alpha} (p_i)\equiv \tilde \lambda_i$ and skip the $SU(8)$ indices on the $\eta$'s.

The essence of the supersymmetric deformation is the invariance of both delta functions, namely the arguments of both delta functions remain $z$-independent.
\be
\sum_i  \lambda_{\alpha i}(z) \tilde \lambda_{\dot \alpha i}(z) = \sum_i  \lambda_{\alpha i} \tilde \lambda_{\dot \alpha i}
\ee
and
\be
\sum_i \tilde \lambda^{\dot \alpha}_{ i}(z) \eta_{i A}(z) = \sum_i \tilde \lambda^{\dot \alpha}_{ i} \eta_{i A}.
\ee
Another useful property of the 2-point shift above is that the triangular and square brackets of these two deformed points,  $k$ and $l$,  are $z$-independent, namely
\be
 \langle k  l \rangle (z) = \langle k  l \rangle \qquad [k l] (z) = [k l].
\ee
 The deformed generating function for the 4-point amplitudes is given by the sum of contribution from all loop orders:
\be
\Omega_4(z) = \delta^4(\sum_i  \lambda_{\alpha i} \tilde \lambda_{\dot \alpha i}) \delta^{16} (\sum_i \tilde \lambda^{\dot \alpha}_{ i} \eta_{i A}) \sum_L  \kappa^{2(L-1)} {\cal P}^L (1^+, 2^+, 3^+, 4^+)(z)
\label{Omega4L} .\ee
In this expression we have used the fact that
as we have shown above, both $\delta$-functions are $z$-independent!
The only $z$ dependent factor is  the function ${\cal P}^L (1^+, 2^+, 3^+, 4^+)$ which is symmetric in all particles,  has +2 helicity at each point and depends on commuting spinors $\lambda$ and $\tilde \lambda$ at each point.

\subsection{Tree level}
 For the tree amplitudes we have
\be
{\cal P}^{L=0} (1^+, 2^+, 3^+, 4^+)= {s_{i, i+1} s_{i, i+3}\over  s_{i, i+2}} \left (\prod_{k=i}^{k=i+4}
 \langle k(k+1)\rangle \right )^{-2} .
\ee
 We may now pick up any two points out of these 4 for the supersymmetric deformation and check that
 \be
{\cal P}^{L=0} (1^+, 2^+, 3^+, 4^+)(z)_{z\rightarrow \infty}  \rightarrow {1\over z^2} .
\ee
Thus the total generating function  at the tree level $\Omega_4^{tree} $ behaves as $1/z^2$. It is easy to preserve this nice large $z$ behavior for the amplitude with two negative helicities. The relation between ${\cal P} (1^+, 2^+, 3^+, 4^+)$ and the amplitude with 2 negative (at $k$ and at $l$ points) and 2 positive graviton helicities involves the $z$-independent factor $ \langle k  l \rangle^8$. In this way good properties of the  function ${\cal P} (1^+, 2^+, 3^+, 4^+)$ lead to good properties of the 2 negative, 2 positive graviton helicity amplitudes, for example
\be
M^{L=0} (1^-, 2^-, 3^+, 4^+)(z)_{z\rightarrow \infty}  \rightarrow {1\over z^2} .
\ee
\subsection{ Deformation of counterterms}

For the 3-loop counterterm appropriate for the helicity formalism as suggested in \cite{Kallosh:2008mq}  we take
 \be
{\cal P}^{3-loop} (1^+, 2^+, 3^+, 4^+)_{UV}=  \kappa^4 {\cal P}^{3}_{UV}  = \left  (  {\kappa^2 s_{12} s_{14} \over
 \langle 1 2\rangle  \langle 23\rangle  \langle 34\rangle  \langle 41\rangle }\right)^2 .
\ee
One can see that for the choice of the  deformation in direction 1 and 2 or in direction 3 and 4  this function is $z$ -independent:
 \be
{\cal P}^{3-loop}_{UV} (1^+, 2^+, 3^+, 4^+)(z)= {\cal P}^{3-loop}_{UV} (1^+, 2^+, 3^+, 4^+) .
\label{largez}\ee
This  means that the (- - + +) 3-loop UV divergent amplitude cannot vanish at large $z$. Moreover, we know that $M^{3-loop}_{UV} \sim  \langle 12 \rangle^4 [34]^4$ and it is simply $z$-independent for the deformation in direction of 1 and 2.
However, it is interesting that in fact the computation in \cite{Bern:2007hh} shows that the total 3-loop counterterm does not appear and therefore the constant in eq. (\ref{largez}) actually vanishes.

Higher loop counterterms for the 4-point function will differ from the 3-loop ones only by multiplication of the polynomials of Mandelstam variables. Therefore the corresponding amplitudes  will also not vanish at large $z$. The conjecture of all-loop finiteness would mean that no such terms would appear in higher-loop computations.

\subsection{Deformation of 1-loop amplitudes}

The  $z$ dependence of the 1-loop UV finite amplitudes is defined by the  $z$ dependence of
 \be
{\cal P}^{1-loop} (1^+, 2^+, 3^+, 4^+)(z)= {\cal P}^{3}_{UV}(z) F^{box}(z) .
\ee
Here ${\cal P}^{3}_{UV}(z)$ is
\be
{\cal P}^{3}_{UV}  (1^+, 2^+, 3^+, 4^+)(z) =   {s_{12} s_{14} \over
 \langle 1 2\rangle  \langle 23\rangle  \langle 34\rangle  \langle 41\rangle } (z) .
\ee
It is $z$-independent for the deformation in 1,2 or 3, 4 directions. The box function for the deformation, say in 1, 2 direction is
\be
F^{box} (z) = -{1\over (-s_{12})^{1+\epsilon} s_{13}(z)}\left [{4\over \epsilon^2} + {2 \ln (s_{12}/s_{13}(z))\over \epsilon} -4\pi^2 /3 \right] + ...
\ee
where ... mean 2 other terms in eq. defining the box function in eq. (\ref{Fbox}).  It  has dimension -8, therefore it depends on ${1\over s_{12} s_{23}}$ as well as on some logarithmic function of Mandelstam variables. When explicit expression for the box function in eq. (4.25) of \cite{Bern:1993kr} or from Appendix B of \cite{Bern:2005iz} is used we find that $F^{box} (z) $
at large $z$  has the following terms
\be
{1\over \epsilon} {\log z\over z} \ , \qquad {\log z\over z} .
\ee
Therefore at large $z$ the one-loop generating function ${\cal P}^{1-loop} (1^+, 2^+, 3^+, 4^+)(z)= {\cal P}^{3}_{UV}(z) F^{box}(z)$ vanishes.

\section{ \E \, symmetry in perturbative amplitudes of \N=8}

It is explained in \cite{KK} that  the unbroken \E \, symmetry  leads to  low-energy theorem which, in principle, may acquire a more general form then just the vanishing of the one-soft scalar limit of the on-shell amplitudes. A {\it schematic form of the low-energy theorem}  which follows from the unbroken \E \, symmetry connects the $n+1$-point amplitude with one soft scalar to the $n$-point amplitude without a scalar:
\be
\partial_\mu J^\mu_{Noether} =0 \quad \Rightarrow \quad {\cal M} (n+\phi(q))_{q \rightarrow 0} =  g_{KU} {\cal M} (n) .
\label{LET1}\ee
A detailed form of these relation is presented in  \cite{Bando:1987br}, \cite{KK}. Here the  Kugo-Uehara (KU) charge  $g_{KU}$ serves as a bridge between the soft $n+1$-point amplitude and the $n$-point amplitude and it is independent on the number of external particles in the amplitude. In some cases it  may vanish. For example,  it was shown to vanish  in \cite{KK} at the tree level in \N=8 \, SG. In such case, the \E \, symmetry, $\partial_\mu J^\mu_{Noether} =0$, is reduced to a simple requirement that the 1-soft-scalar limit vanishes:
\be
 {\cal M}^{tree} (n+\phi(q))_{q \rightarrow 0} =  0 \ ,    \qquad g_{KU}^{tree}=0 .
\ee
However, in general case the symmetry in higher loops may be valid due to the cancelation between the two terms in the low-energy theorem:
\be
  \langle n-m | \partial_\mu  J^\mu_{Noether} | m \rangle = {\cal M} (n+\phi(q))_{q \rightarrow 0} - g_{KU} {\cal M} (n)=0 .
\label{LET}\ee
 The symmetry only requires that the  limit $q \rightarrow 0$ of the $n+1$-point amplitude is related in a certain way to the $n$-point amplitude without a scalar.  Therefore in a generic situation beyond tree level, the test of the \E \,  symmetry includes three  steps. First, one has to evaluate the one-soft scalar limit of the on-shell $n+1$-point amplitude.   Secondly, one has to evaluate the second term in the low-energy theorem proportional to the $n$-point amplitude without a scalar, via the KU charge. The third point is to  check if the two terms in eq.  (\ref{LET})  cancel. In such case the  \E \, symmetry  has no anomalies since the current is conserved with account of quantum corrections.

Without this additional study,  which corresponds to the computation of the KU charge  via the matrix elements of the Noether current \cite{Kallosh:2008ic} between 1-particle states, one cannot make a definite statement about the \E \, symmetry on the basis of the vanishing of the 1-soft-scalar limit of the $n+1$-point  amplitude. In the tree approximation where no anomalies are expected, $\partial_\mu J^\mu_{Noether}=0$ is taken for granted. To confirm the expected symmetry it was necessary to compute both terms in eq.   (\ref{LET1}).  The left hand side in  eq. (\ref{LET1}),  ${\cal M} (n+\phi(q))_{q \rightarrow 0}$,  was computed in   \cite{Bianchi:2008pu} and in \cite{ArkaniHamed:2008gz}, and was found to vanish.  The term on the right hand side,   $g_{KU} {\cal M} (n)$  was computed in \cite{Kallosh:2008ic} and was found  to vanish. These two computations confirm the unbroken symmetry and the low-energy theorem where each of the two terms vanishes separately.

  The case of the 4-point amplitudes in \N=8 \, SG is special. If we assume that the \E \, symmetry is unbroken it means that the soft limit of the 4-point function is related to the 3-point function via the KU charge
  \be
  \langle 3-m | \partial_\mu J^\mu_{Noether} |m \rangle= {\cal M} (3+\phi(q))_{q \rightarrow 0} - g_{KU} {\cal M} (3)=0 .
\label{LET3}\ee
   In \N=8 \, SG the on-shell 3-point function  vanishes for real momenta, ${\cal M} (3)=0$. This means that the requirement of unbroken \E \, symmetry is equivalent to the requirement that the 4-point amplitude has a vanishing soft limit, independently of the value of the KU charge, ${\cal M} (3+\phi(q))_{q \rightarrow 0} =0$.
Here we have explicit expressions for the 4-point scalar  as well as 2-scalar-2-vector amplitudes and we may check the \E \, symmetry of the 3-loop counterterm as well as 1-loop amplitudes.   If the limit vanishes for real momenta, it means that the Noether current \cite{Kallosh:2008ic} is  conserved and \E \, symmetry is anomaly-free. We will consider  the following limit:
\be
p_3\sim \delta^2  \ ,  \qquad \tilde \lambda_3\sim  \lambda_3 \sim \delta \ , \qquad s\sim t \sim u \sim \delta^2 \ , \qquad \delta \rightarrow 0 .
\ee
We may also perform a more interesting analysis of the soft limit of the 4-particle amplitudes at complex momenta, in the spirit of  \cite{ArkaniHamed:2008gz}.   In such case the 3-point amplitude is not vanishing, ${\cal M} (3)\neq 0$.  We assume that eq.  (\ref{LET3}) remains valid for the complex momenta. In terms of commuting spinors  we will consider the soft limit at complex momenta as follows
\be
\tilde \lambda_i \neq \lambda^*_i \ , \qquad \lambda_3 \rightarrow 0 \quad \Rightarrow \quad p_3 \rightarrow 0 .
\ee
Since  $p_i^2=0$ and $\sum_i p_i =0$ we keep the relation between Mandelstam variables $s+t+u=0$ and
\be
-s = 2p_1 p_2= 2p_3p_4 \, \qquad -u = 2p_1 p_4= 2p_2 p_3 \ , \qquad -t = 2p_1 p_3= 2p_2 p_4 .
\ee
Each of the Mandelstam variables is  linear in $\lambda_3$.  Also linear in $\lambda_3$ are the square brackets $[i3]=  \lambda_{i }^{ \alpha}  \lambda_{3  \alpha}$ with $i=1,2, 4$ (we use notation from \cite{Bianchi:2008pu}).  The triangular brackets $ \langle ij \rangle = \tilde \lambda_{i \dot \alpha} \tilde \lambda_j ^{\dot \alpha} $ are not small.
We will also study the  limit of small $p_3$  via the soft  $\tilde \lambda_3$
\be
\tilde \lambda_i \neq \lambda^*_i \ , \qquad \tilde \lambda_3 \rightarrow 0 \quad \Rightarrow \quad p_3 \rightarrow 0 .
\ee
In this case the triangular brackets $\langle i3 \rangle $ are small whereas the square ones are not small.

\subsection{Tree level 4-point amplitudes}

The study of the soft amplitudes in general can be simplified if we first confirm the \E \, symmetry of the tree 4-point amplitudes which states that  ${\cal M} (3+\phi(q))_{q \rightarrow 0} =0$.
Here we will first look at the  2-scalar-2-vector tree amplitudes in eq. (\ref{2+2tree}) where we have to check the limit when one of the scalar momenta vanishes, for example $p_3\rightarrow 0$.
There are 3 type of terms
\be
 \frac{\langle13\rangle^2 [23]^2 }{s} \ ,  \qquad  \frac{\langle13\rangle^2 [23]^2 }{t} \ , \qquad \frac{\langle13\rangle^2   [23]^2}{u} .
 \label{2+2tree1}\ee
 For the 4-scalar tree amplitude in eq.  (\ref{eqn:4scalars})
 we have the following structures
 \be
 {tu\over s} \ ,\qquad {su\over t} \ , \qquad {ts\over u} \, \qquad s \ , \qquad t \ , \qquad u .
 \ee
In the real limit, $\tilde \lambda_3\sim \delta$,  $\lambda_3 \sim \delta$ and  $s\sim t \sim u \sim \delta^2$ each of the amplitudes behaves as $\delta^2$.
For complex momenta we take $\lambda_3 \sim \delta$ and $p_3\sim s\sim t\sim u\sim \delta$,  or  $\tilde \lambda\sim \delta$ and $p_3\sim s\sim t\sim u\sim \delta$.
Each amplitude  is clearly vanishing as $\delta$ when either  $\lambda$ or  $\tilde \lambda$   vanishes.
If we instead decide to take a limit  $p_4\rightarrow 0$, we have to take into account that $\langle13\rangle  [23] = - \langle14\rangle [24]$ and we get the same result, namely the tree 2-scalar-2-vector amplitude vanishes in any of the  soft scalar limit which we study.

 \subsection{3-loop UV divergent amplitudes}

 The candidate for the 3-loop divergence  (\ref{superaction}) was constructed only at the linear level. The  linearized \E \, symmetry of it was partially established in \cite{Howe:1981xy} for the 4-scalar amplitudes.  At the linear level \E \, symmetry  is reduced to a shift symmetry of scalars since  the 3-point counterterm is vanishing. This  means that when any of the scalars in the 4-point amplitude has a soft momentum $q$, the amplitude should vanish in the soft limit  $q\rightarrow 0$.

 The structure of 4-point 3-loop counterterm  presented in Section 6 differs from the tree amplitudes  by a factor $stu$. For the
 2-scalar-2-vector  amplitudes we get the following terms
\be
\langle13\rangle^2 [23]^2 tu \ ,  \qquad  \langle13\rangle^2 [23]^2 su \ , \qquad \langle13\rangle^2   [23]^2 st .
 \label{2+23loop}\ee
 All these terms behave as $\delta^8$ in the real soft limit and as
 as $\delta^4$  in case of small $\lambda_3$ as well as in case of small $\tilde \lambda_3$. For the 4-scalar  amplitude
  \be
 t^2u^2 \ ,\qquad s^2u^2 \ , \qquad t^2s^2 \, \qquad s^2tu \ , \qquad t^2su \ , \qquad u^2st
 \ee
the situation is analogous.
 Therefore in the one-soft-scalar limit the counterterm amplitudes behave  softer than the tree amplitudes. This means that   the linearized \E \, symmetry of the 3-loop counterterm is established.  This suggests that the 3-loop finiteness is not due to the absence of the  linearized \E \, symmetry of the 3-loop counterterm and we need other explanation of the 3-loop finiteness.

  \subsection{1-loop UV finite amplitudes with scalars}

We are interested in 4-point amplitudes with scalars at the 1-loop level since we would like to test the linearized \E \, symmetry.
The generic formula follows from the relation discussed before, $
{\cal P}^{1-loop} (1^+, 2^+, 3^+, 4^+)= {\cal P}^{3}_{UV} F^{box}
$ where $F^{box}$ is defined in eq. (\ref{Fbox}). An explicit expressions for the 1-loop  amplitude for 2 vectors and 2 scalars is
\bea
&&M_{2vec2sc}^{1-loop}( b_{AB}^-, b_+^{CD}, b^{EFGH}, b^{IJKL} ) = - \langle13\rangle^2 [23]^2 \times \nonumber\\
\nonumber\\
&&\left[ t\, u\, \delta^{CD}_{AB}\epsilon^{EFGHIJKL} + 3! \, s\, u\, \delta_{AB}^{[EF }\epsilon^{GH]IJKLCD}+3! \, s\, t\, \delta_{AB}^{[IJ}\epsilon^{KL]EFGHCD}\right]  F^{box} .\label{1loop2+2}\eea
The 1-loop 4 scalar amplitude is
\begin{align}
&M_{4sc}^{1-loop} ( b^{ABCD},  b^{EFGH}, b^{IJKL}, b^{MNPQ}\rangle=\cr
&(t^2\, u^2\, \epsilon^{ABCDEFGH}\epsilon^{IJKLMNPQ}+s^2\, u^2\, \epsilon^{ABCDIJKL}\epsilon^{EFGHMNPQ}
+s^2 \, t^2\, \epsilon^{ABCDMNPQ}\epsilon^{EFGHIJKL})  F^{box}\cr
&+\frac{1}{2(4!)^3}\sum_{\textit{perm}}(-1)^\textit{perm}\left[s^2\, t\, u\, \epsilon^{1_11_21_31_43_13_24_34_4}\epsilon^{2_12_22_32_44_14_23_33_4}+ s\, t^2\, u\, \epsilon^{1_11_21_31_42_12_24_34_4}\epsilon^{3_13_23_33_44_14_22_32_4}+\right.\cr
&\left.s\, t\, u^2\,  \epsilon^{1_11_21_31_42_12_23_33_4}\epsilon^{4_14_24_34_43_13_22_32_4}\right] F^{box} .
\label{1loop4scalars}
\end{align}

Thus, for the 2-scalar-2-vector case and 4 -scalar case of 1-loop UV finite amplitudes we  get the following dependence on momenta
  \be
\langle13\rangle^2 [23]^2 tu \;  F^{box}  \ ,  \qquad  \langle13\rangle^2 [23]^2 su \; F^{box}\ , \qquad \langle13\rangle^2   [23]^2 st \; F^{box}
 \label{2+21loop}\ee
  and
    \be
 t^2u^2  \;  F^{box} \ ,\qquad s^2u^2  \;  F^{box} \ , \qquad t^2s^2  \;  F^{box} \, \qquad s^2tu   \;  F^{box} \ , \qquad t^2su  \;  F^{box} \ , \qquad u^2st  \;  F^{box} .
 \ee
 All these terms in the soft limit behave as  $\delta^8 F^{box}(\delta)$ in the real case and as
  $\delta^4 F^{box}(\delta)$ for either $\tilde \lambda_3\sim \delta$ or  $\lambda_3 \sim \delta$ . The $ F^{box}$  function, according to our dimensional analysis has dimension - 8 and up to some logarithmic functions depends on Mandelstam variables as follows
  \be
 F^{box}\sim {f_0\over st}  +{g_0\over tu} + {e_0\over us} .
  \ee
  Here the functions $f_0, g_0, e_0$  have dimension zero and may depend on some logarithmic functions of momenta. For the soft limit this will not be relevant since $F^{box} \sim \delta^{-4}$ up to some logarithmic dependence on $\delta$ in the real case and as  $F^{box} \sim \delta^{-2}$  for soft spinors of only one kind. The total 1-loop amplitudes therefore in the soft limit inherit  8 (4) powers of softness from the 3-loop amplitudes and -4 (-2)  powers from the box for the real (complex) case. The result is $\delta^4$ up to logarithmic terms for the real limit and  $\delta^2$ up to logarithmic terms for the real limit. This  proves  the linearized \E \, symmetry of the 1-loop 4-point amplitudes.

  \subsection {$n+1$-point 1-loop amplitudes with $n>3$}

  The $n+1$-point 1-loop amplitudes with $n>3$ with scalars are not available in the explicit form. However, on the basis of the information above which we have got from the case of the 4-point amplitudes, we can clarify the relation between the  \E \, symmetry and the 1-soft-scalar limit of the $n+1$-point 1-loop amplitudes with $n>3$.  For this purpose we
 will  {\it assume that  $\partial_\mu J^\mu_{Noether}=0 $ is also valid for the complex momenta} for the 4-point amplitudes and relay on eq. (\ref{LET3}) for complex momenta.

  We have found above that the soft limit of the 4-point amplitude is vanishing for complex momenta. Since  the 3-point amplitude is not vanishing in this case, from eq.  (\ref{LET3}) we deduce that the KU charge is vanishing for  all 1-loop amplitudes, $g_{KU}^{1-loop}=0$. Now we can take into account that  this charge is a universal bridge between the $n+1$ amplitude with a soft scalar and an $n$-point amplitudes with all hard momenta and use this information in eq. (\ref{LET}). This  provides us with  the simple form of low-energy theorem for the unbroken \E \, symmetry for the 1-loop $n+1$-point amplitudes  with $n>3$:
   \be
  \langle n-m | \partial_\mu  J^\mu_{Noether} | m \rangle ^{1-loop} = {\cal M}^{1-loop} (n+\phi(q))_{q \rightarrow 0} =0 \qquad n>3  \qquad \rm since \qquad g_{KU}^{1-loop}=0 .
\label{LET+3}
\ee
It remains to find out if ${\cal M} (n+\phi(q))_{q \rightarrow 0}=0$ for $n>3$ to complete the test the  \E \, symmetry of all 1-loop amplitudes.

\

 \section{Discussion}

 In \N=8 \, SG the supersymmetry guarantees that there are no divergences at 1  or 2 loops, and it was expected for many years that the first divergence would appear at 3 loops. The calculation has now been done \cite{Bern:2007hh}, and
there is no 3-loop divergence.  This raises the possibility that these cancelations continue to higher orders, and that the amplitudes are UV finite order by order.

During the last few years there was a remarkable progress in computational tools of the amplitudes in \N=4 \, YM and \N=8 \, SG
which was not based on the Feynman path integral but on unitarity cut method, on recursion relations, complex deformation of momenta, helicity formalism etc.  In this paper we were trying to use these new methods  to understand better the properties of the counterterms, the 3-loop one in the first place. We also studied  the restrictions which the \E \, symmetry puts on the on-shell perturbative amplitudes of \N=8 \, SG.

We have computed the  4-point scalar amplitudes, candidates for the 3-loop UV divergence, in the helicity formalism and tested their \E \, symmetry. We prepared them in the form in which they can be compared with the light-cone counterterms. We also developed the supersymmetric deformation of the 4-point generating function and studied the large complex momentum behavior of the counterterms.  Using the explicit form of the 1-loop 4-point scalar amplitudes derived in this paper we have verified that the continuous \E \, symmetry is respected by the quantum corrections of \N=8 \, SG.

Our findings so far are in agreement with the fact that the $SU(8) $ symmetry has no anomalies at the  1-loop level \cite{Marcus:1985yy}. It has been argued in \cite{Kallosh:2008mq} that this implies the absence of \E \, anomalies. Our explicit computation of the 1-loop 4-point scalar  amplitudes confirms the linearized form of the  \E \, symmetry and absence of anomalies. With account of the vanishing soft limit for complex momenta of the 1-loop 4-point scalar  amplitudes  we derived   the low-energy theorem (\ref{LET+3}) for the 1-loop $n$-point amplitudes which is simpler than the generic case (\ref{LET}) and  which may be useful for the future studies. In particular, if we believe that  the absence of 1-loop $SU(8)$ anomalies is an indication of the absence of the \E \, anomalies, our low-energy theorem (\ref{LET+3})  predicts that all 1-soft-scalar limit of $n$-point amplitude should vanish.

We hope that the results in this paper may help to  pursue the next levels of investigations of the UV structure of \N=8 \, SG.

\section*{Acknowledgments}

We are grateful to  L.~Alvarez-Gaume, N. Arkani-Hamed,   M.~Bianchi, F. Cachazo,   H.~Elvang, L.  Dixon,  S. Ferrara, D. Forde, D. Freedman,  J.~Kaplan, G. Korchemsky, T. Kugo, A. Linde, S. Shenker,  E. Sokatchev,  M. Soroush, K. Stelle and L. Susskind for the most useful discussions of \N=8 supergravity.
 This work  was supported by the NSF grant~ 0756174.

\end{document}